\documentclass[%
 aip,
 amsmath,amssymb,
 reprint,%
]{revtex4-1}

\usepackage{graphicx}
\usepackage{dcolumn}
\usepackage{bm}

\usepackage[utf8]{inputenc}
\usepackage[T1]{fontenc}
\usepackage{mathptmx}
\usepackage{etoolbox}

\usepackage{color}
\usepackage{float}
\usepackage[hidelinks=true]{hyperref}
\hypersetup{
  colorlinks   = true, 
  urlcolor     = blue, 
  linkcolor    = blue, 
  citecolor   = red 
}

\makeatletter
\def\@email#1#2{%
 \endgroup
 \patchcmd{\titleblock@produce}
  {\frontmatter@RRAPformat}
  {\frontmatter@RRAPformat{\produce@RRAP{*#1\href{mailto:#2}{#2}}}\frontmatter@RRAPformat}
  {}{}
}%
\makeatother
\begin{document}

\preprint{AIP/123-QED}

\title{Generalized  Bond Polarizability model for more accurate atomistic modeling of Raman spectra}
\author{Atanu Paul}
 \affiliation{Department of Chemistry, Bar-Ilan University, Ramat Gan 5290002, Israel}
\author{Nagaprasad Reddy Samala}
 \affiliation{Department of Chemistry, Bar-Ilan University, Ramat Gan 5290002, Israel}
\author{Ilya Grinberg}
\email[]{ilya.grinberg@biu.ac.il}
\affiliation{%
Department of Chemistry, Bar-Ilan University, Ramat Gan 5290002, Israel
}%


\begin{abstract}
Raman spectroscopy is an important tool for studies of molecules, liquids and solids. While Raman spectra can be obtained theoretically from molecular dynamics (MD) simulations, this requires the calculation of the electronic polarizability along the simulation trajectory.  First-principles calculations of electronic polarizability are computationally expensive, motivating the development of atomistic models for the evaluation of the changes in the electronic polarizability with the changes in the atomic coordinates of the system.  The bond polarizability model (BPM) is one of the oldest and simplest such atomistic models, but cannot reproduce the effects of angular vibrations, leading to inaccurate modeling of Raman spectra.  Here, we demonstrate that the generalization of BPM through inclusion of terms for atom pairs that are traditionally considered to be not involved in bonding dramatically improves the accuracy of polarizability modeling and Raman spectra calculations. The generalized BPM (GBPM) reproduces the $ab$ $initio$ polarizability and Raman spectra for a range of tested molecules (SO$_{2}$, H$_{2}$S, H$_{2}$O, NH$_{3}$, CH$_{4}$, CH$_{3}$OH and CH$_{3}$CH$_{2}$OH) with high accuracy and also shows significantly improved agreement with $ab$ $initio$ results for the more complex ferroelectric BaTiO$_{3}$ systems.
For liquid water, the anisotropic Raman spectrum derived from atomistic MD simulations using GBPM evaluation of polarizability shows significantly improved agreement with the experimental spectrum compared to the spectrum derived using BPM.
Thus, GBPM can be used for the modeling of Raman spectra  using large-scale molecular dynamics and provides a good basis for the further development of atomistic polarizability models.          
\end{abstract}

\maketitle

\section{INTRODUCTION}
Since the discovery of the Raman effect by C. V. Raman, measurement of Raman spectra has become an essential tool for understanding  various structural and vibrational properties of molecules and solids.~\cite{chemosensors9090262,long,PhysRevLett.118.136001} For instance, Raman spectroscopy has been  used to study materials properties under strain,~\cite{Wolf_1996} different types of defects~\cite{acsnano.5b07388,PhysRevB.76.024303} and for the study of ferroelectric domain walls.~\cite{Rubio-Marcos2015}
However, accurate interpretation of the peaks of Raman spectra is still a difficult and challenging task.  

Theoretical calculation of Raman spectra based on molecular dynamics (MD) simulations requires the trajectory of the system electronic polarizability. Raman spectra are then obtained through  the Fourier transform of the autocorrelation function of the polarizability trajectory.~\cite{Raman_intensity} In contrast to the first-principles calculations of the Raman modes and frequencies, this approach takes into account the anharmonicity of the potential energy surface and  thermal broadening effects,  and can obtain the intensities as well as the positions of the Raman peaks.~\cite{PhysRevLett.88.176401}  Calculation of the polarizability trajectory at each time step using density functional perturbation theory (DFPT)~\cite{DFPT_baroni} is very expensive and almost impossible for large systems (e.g. 10,000-100,000 atoms) and long simulation trajectories ($> 1$~ns) even with modern computational resources. This has stimulated research in atomistic modeling of the  electronic polarizability changes during molecular dynamics simulations. The atomistic polarizability models are parameterized either using  the  polarizability values obtained  from $ab$ $initio$ calculations or by fitting to experimental Raman spectra. In this approach, the simplest and oldest atomistic model is the bond-polarizability model (BPM),~\cite{BPM1,1996273} which has been successfully applied to various systems for the calculations of Raman spectra.~\cite{C7NR05839J,PhysRevB_241402,Luo2015,PhysRevB.63.094305} However, as BPM is based on the changes of the  bond lengths only and completely neglects the importance of bond-bond interactions, it can also fail to reproduce the DFPT or experimental results in some cases.~\cite{ghosez,berger2024polarizability,atanu_BPM} In an alternative approach, Thole model was introduced to calculate molecular polarizability based on dipole interaction,~\cite{THOLE1981341,Thole_1} and an improved representation of the Thole model (TholeL) that considers the bond-dependent dipole interaction was also introduced to calculate molecular polarizabilities.~\cite{TholeL}

In addition to simple models, molecular polarizability was also simulated using more complex polynomial expansion and machine learning (ML) models.  A full polynomial expansion~\cite{polynomial} of the polarizability tensor involves a  large  number of  parameters; for instance, 128 independent parameters were used to model the molecular polarizability of H$_{2}$O.~\cite{polynomial} Recently, ML-based polarizability model have become popular and have been applied to molecules, liquids and solids.~\cite{LUSSIER2020115796,car,berger2024raman_N,Raimbault_2019,acs.jpcc.4c00886}  However, polarizability models based on the polynomial expansion and machine learning requires a large  training data set  generated from expensive DFPT calculations to determine the unknown parameters. Furthermore, such models are much more computationally expensive than simple atomistic models, and thus are less suitable for use in large-scale molecular dynamics simulations.

To address the need for simple yet accurate atomistic models of electronic polarizability, here we introduce an improved bond polarizability model that is based on the generalization of the bond polarizability concept to atom pairs that are traditionally considered not to be involved in chemical bonding. We show that such generalized bond polarizability model (GBPM) reproduces the DFPT-calculated electronic polarizability for simple molecules (H$_{2}$O, SO$_{2}$, H$_{2}$S, NH$_{3}$, CH$_{4}$, CH$_{3}$OH, CH$_{3}$CH$_{2}$OH) and achieves a significant improvement in accuracy for the more complex solid-state BaTiO$_{3}$ (BTO) and liquid water systems, making GBPM suitable for obtaining Raman spectra  from MD simulations and providing a good basis for the further development of accurate atomistic models of electronic polarizability.


The rest of the paper is organized as follows. We first introduce the concept of GBPM and describe our computational methodology in section II. We then present our results and discussion of the considered systems in section III. Finally, we draw the conclusions in section IV.  

\begin{figure}[]
\centering
\includegraphics[width = 8.5 cm,angle =0]{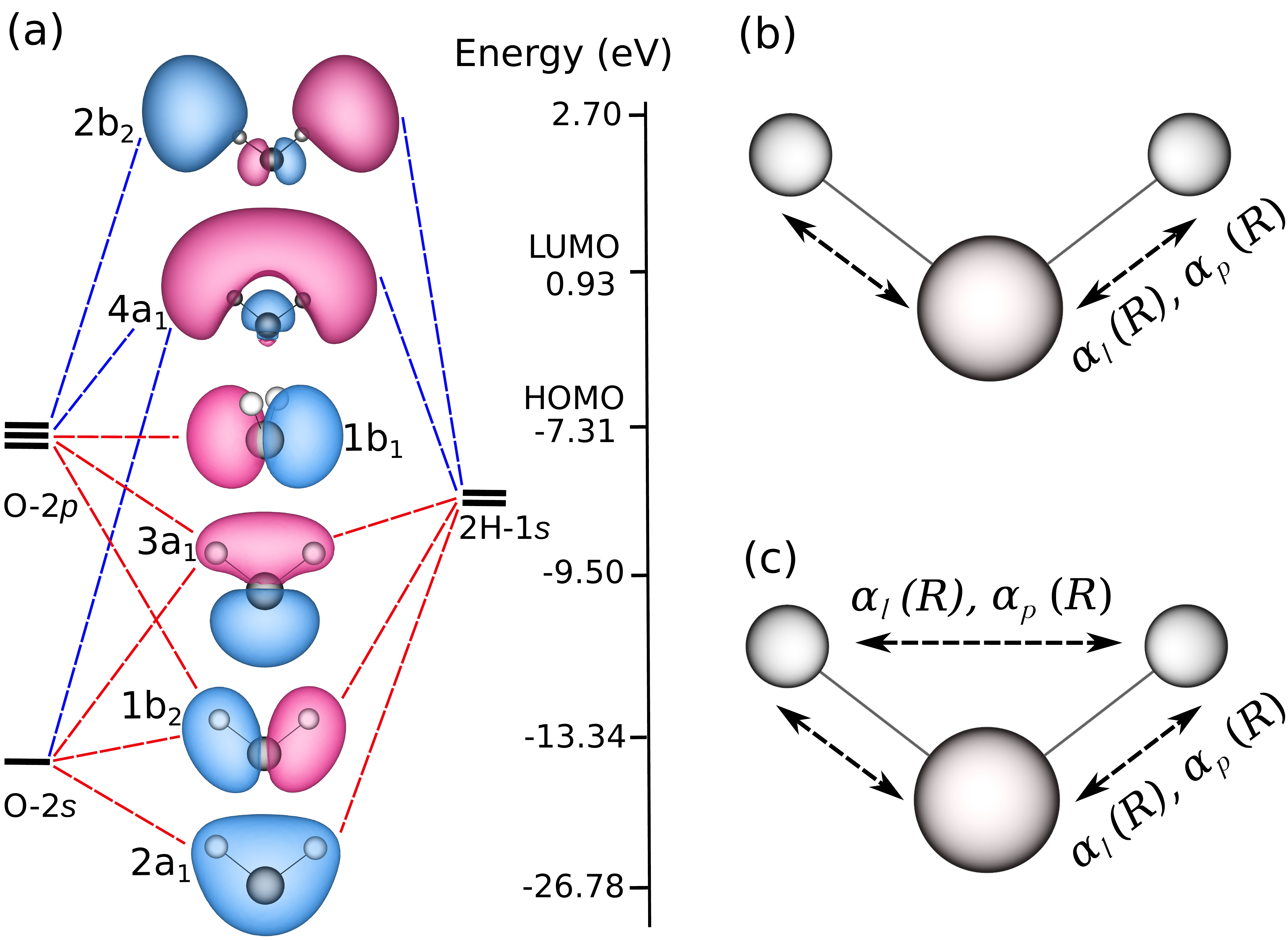}
\caption{(a) Molecular orbitals of H$_{2}$O labeled with respective symmetries and energies calculated in Gaussian~\cite{g16} code using PBE functional.~\cite{PhysRevLett.77.3865} Schematic representation of (b) BPM, (c) GBPM. $\alpha_{l}$ and $\alpha_{p}$ are parallel and perpendicular bond-polarizability, respectively.}
\label{Fig1} 
\end{figure}

\section{Theory and COMPUTATIONAL METHODS}

\begin{figure*}[]
\centering
\includegraphics[width = 16.0 cm,angle =0]{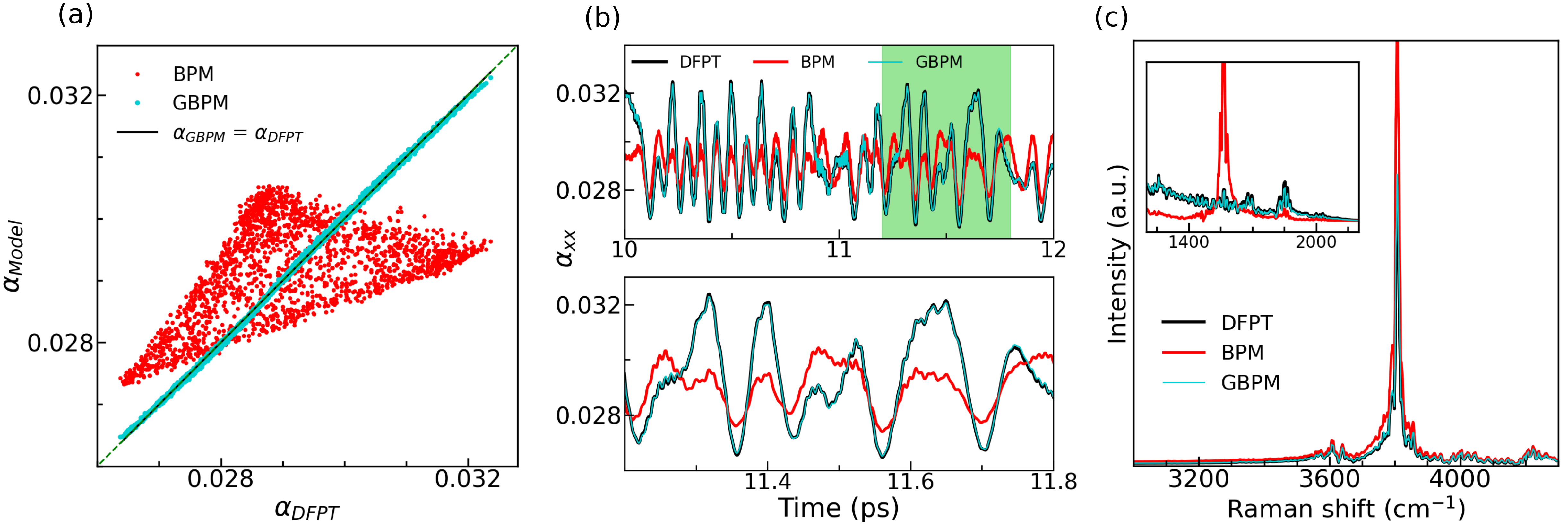}
\caption{H$_{2}$O (a) $xx$ component of $\alpha_{BPM}$ (red circle) and $\alpha_{GBPM}$ (cyan circle) order vs $\alpha_{DFPT}$. Linear fit of GBPM order is shown in solid black line. Green dotted line represents $\alpha_{Model}$ = $\alpha_{DFPT}$. (b) Top panel: Comparison of the trajectory of $\alpha_{xx}$ from DFPT (black), $\alpha_{BPM}$ (red) and $\alpha_{GBPM}$ (cyan) order. Bottom Panel: Magnified view of the region shaded in filled green in the first panel. (c) Raman spectra from DFPT (black), BPM (red) and GBPM (cyan). Inset shows the results in the lower-wavenumber region.}
\label{Fig2} 
\end{figure*}

In the BPM, the total polarizability ($\alpha_{ij}$) of the system of volume ($\Omega$) can be obtained by summing over the polarizability contributions of the  bonds. Therefore, we have 
\begin{equation}
\alpha_{ij} = \frac{1}{\Omega} \sum_{n} \alpha_{ij}^{n}
\end{equation}
where $i$, $j$ denote the Cartesian axes. $\alpha_{ij}^{n}$ is the polarizability contribution of the $n^{th}$ bond. Each bond polarizability is further defined in terms of parallel and perpendicular contributions as,
\begin{equation}
\alpha_{ij}^{n} = \frac{1}{3}(\alpha_{l} + 2\alpha_{p})\delta_{ij} + (\alpha_{l} - \alpha_{p})(\frac{R_{i}R_{j}}{R^{2}} - \frac{1}{3}\delta_{ij}) 
\end{equation}
where \pmb{$R$}, $\alpha_{l}$ and $\alpha_{p}$ are the $n^{th}$ bond vector, longitudinal and perpendicular bond polarizability, respectively. Furthermore, $\alpha_{l}$ and $\alpha_{p}$ depend only on the bond length ($R$) and therefore can be expanded in the Taylor series around the equilibrium bond length ($R^{0}$),
\begin{equation}
\alpha_{t} = \alpha_{t}^{0} + \alpha_{t}^{1} (R - R^{0}) + \alpha_{t}^{2} (R - R^{0})^{2} + .. 
\end{equation}
where $\alpha_{t}$ is either longitudinal ($t$ = $l$) or perpendicular ($t$ = $p$) component of the bond polarizability. $\alpha_{t}^{0}$, $\alpha_{t}^{1}$, $\alpha_{t}^{2}$ are constants for a particular bond type. Fig.~\ref{Fig1} (b) shows a schematic representation of the BPM where total polarizability of the nonlinear triatomic molecule is calculated in terms of bond polarizability contributed by the two bonds marked in the figure.

To improve on the BPM and include the effects of angular vibration, we reconsider the fundamental assumption of BPM which is that the polarizability is due to the variations in the distances of the bonds between the atoms in the molecules. This equivalent to assuming  that the charge density of the molecule exists only along the bonds or that the response of the charge density to the electric field is dominated by the  charge density along the chemical bonds.  While chemically intuitive, this assumption is incorrect as can be seen by the examination of the molecular orbitals of H$_{2}$O, where a significant charge density is present between the two H atoms for the 2a1 and  3a1 occupied orbitals (see Fig.~\ref{Fig1} (a)).  Additionally for the  1b1 occupied orbital, while there is zero charge density along the vector connecting two H atoms (\pmb{$R$}$_{HH}$)
, a substantial charge density exists along the planes parallel to \pmb{$R$}$_{HH}$.
Thus, in addition to the charge density along the O-H bonds, the charge density along (or parallel to)  \pmb{$R$}$_{HH}$  will also participate in the response to the applied electric field and its response will be affected by angular vibrations that  change \pmb{$R$}$_{HH}$ while leaving  the two O-H distances unchanged.

Furthermore, we consider the quantum mechanical expression for atomic polarizability derived from second-order perturbation theory,
\begin{equation}
\alpha_{ij} \propto \sum_{\gamma\neq 0} \frac{\langle\Phi_{0}\mid\hat{p_{i}}\mid\Phi_{\gamma} \rangle \langle\Phi_{\gamma}\mid\hat{p_{j}}\mid\Phi_{0}\rangle}{\hbar(\omega_{\gamma} - \omega_{0})}
\end{equation}
where $\mid\Phi_{0} \rangle$ and $\mid\Phi_{\gamma} \rangle$ are ground and excited states with energies $\hbar$$\omega_{0}$ and $\hbar$$\omega_{\gamma}$, respectively.
$\hat{p_{i}}$ is the induced electric dipole operator. From this expression, it is clear that frontier orbitals are likely to dominate the polarizability.
Among the unoccupied orbitals, the 4a1 orbital,  which is the LUMO, has a fairly high density in the region between the two H atoms. Therefore, the high charge density of the 4a1 orbital parallel to \pmb{$R$}$_{HH}$
will make the polarizability sensitive to the changes in the H-H distance through the effect of this change on the LUMO energy and wavefunction and the transition matrix elements that set the strength of the electron cloud response to the applied electric field. Similarly, the high charge density along and parallel to \pmb{$R$}$_{HH}$ for the 3a1 HOMO-1 and 1b1  HOMO, respectively, will make the response of the electron cloud  to the applied  electric field sensitive to the changes in the H-H distance.
Thus, the neglect of the H-H distance in BPM is likely to lead to a large error in modeling the variation of $\alpha_{ij}$ during the MD simulation. Therefore, we generalize the BPM to include terms even for the atom pairs that are considered to be non-bonded.  While BPM considers atom pairs connected by one bond, the generalized BPM (GBPM) includes terms for distances between atoms separated by two bonds, thus including the effects of angular vibrations.  We treat the additional pair interactions in exactly the same manner as the bond atom pairs and use second-order Taylor expansion to express their contribution to $\alpha$ as shown in a schematic representation of GBPM in Fig.~\ref{Fig1} (c).
We note that since the GBPM approach is based on the Taylor expansion around the equilibrium bond length, it is unsuited for treating  highly stretched bonds or weak  bonds that undergo large fluctuations in bond lengths such as hydrogen bonds. A more general approach must be developed to take into account the effects of hydrogen bonds on molecular polarizability. However, this is out of the scope of the present paper.


To determine the  parameters of BPM and GBPM, we considered the polarizability trajectory calculated from DFPT~\cite{DFPT_baroni} for the trajectory of $ab$ $initio$ Born-Oppenheimer molecular
dynamics simulations performed using the Quantum ESPRESSO code.~\cite{QE} Optimized norm-conserving pseudopotentials (ONCV)~\cite{ONCV,PseudoDojo} with an energy cutoff of 50 Ry were used employing the Perdew-Burke-Ernzerhof exchange-correlation functional for solids (PBEsol)~\cite{PBEsol} for all the systems considered in this study. The equilibrium structures of all the systems are considered to be the relaxed structure obtained from DFT. The model parameters were obtained by minimizing the difference between the diagonal elements of the model-polarizability tensor and the DFPT-calculated polarizability tensor for a few 
training data set along the DFPT-calculated polarizability trajectory. In our study we consider upto 2$^{nd}$ order term of the Taylor series expansion which will introduce six unknown constants for each type of bond.

For all  molecules (SO$_{2}$, H$_{2}$S, H$_{2}$O, NH$_{3}$, CH$_{4}$, CH$_{3}$OH, CH$_{3}$CH$_{2}$OH) considered in this study, $ab$ $initio$ molecular dynamics were performed at 300~K with a time step of 0.0015 ps. DFPT-calculated polarizability trajectories were obtained for 24, 24, 25 and 24 ps with a time step of 0.0015 ps for SO$_{2}$, H$_{2}$S, H$_{2}$O and NH$_{3}$, respectively. However, for larger molecular systems such as
CH$_{4}$, CH$_{3}$OH and CH$_{3}$CH$_{2}$OH, we considered the DFPT-calculated polarizability trajectories for 23, 23 and 18 ps with the time steps of 0.0058, 0.0058 and 0.0044 ps, respectively, to reduce the computational cost.  For BTO, $ab$ $initio$ molecular dynamics were performed on a 10-atom cell
at four temperatures of 300 K, 600 K, 800 K and 1000 K
for 30 ps with a time step of 0.0015 ps while keeping the volume fixed during the dynamics.
For all of these temperatures, the polarizability trajectory data  used to determine the  BPM and GBPM parameters were obtained from DFPT calculations of the structures along the 30-ps molecular dynamics trajectory sampled at a time step of 0.073 ps.
~For liquid water, classical molecular dynamics were performed using the  LAMMPS code~\cite{LAMMPS} and the  AMOEBA force fields~\cite{AMOEBA1,AMOEBA2,AMOEBA3} with a time step 0.2 fs in a cell of volume 18.64$^{3}$ \AA$^{3}$ containing 216 water molecules. To calculate the Raman spectra of liquid water, we considered a 100-ps molecular dynamics trajectory and the polarizability was calculated at a time step of 0.003 ps.

To calculate the water
  Raman spectra, we considered the time derivative of the polarizability trajectory instead of direct polarizability trajectory with the spectra is obtained by performing the Fourier transform of the autocorrelation function of the polarizability time derivative.~\cite{Raman_intensity} For the molecules and the BaTiO$_3$ solid, we compared the Raman spectra corrsponding to the diagonal compomonent of the polarizability matrix obtained by the BPM and GBPM models with the corresponding spectra obtained from the results of DFPT polaizability calculation.  This is because comparison of the Raman spectra derived from the individual component of the polarizability matrix provides a better way to compare the fine features of the calculated spectra.~\cite{atanu_BPM} For liquid water, the intensity ($R_{\textnormal{aniso}}(\omega)$) of the reduced anisotropic Raman spectra of liquid water was calculated for comparison with the experimental  reduced anisotropic Raman spectrum using the following expression,~\cite{car,Iuchi,Bursulaya}
\begin{equation}
R_{\textnormal{aniso}}(\omega) = n_{\textnormal{BE}}(\omega)\int_{-\infty}^{\infty}\textnormal{d}t \;\;\; \textnormal{cos}(\omega t) \;\; \textnormal{Tr}\textlangle \pmb{\dot{\beta}}(0).\pmb{\dot{\beta}}(t)\textrangle 
\end{equation}

where $\pmb{\beta}$ is the anisotropic part of the polarizability tensor ($\pmb{\alpha}$) 
and $\pmb{\beta}$ can be obtained after subtracting the isotropic part $\overline{\pmb{\alpha}}$ ($\overline{\pmb{\alpha}}$ = $\frac{1}{3}\textnormal{Tr}(\pmb{\alpha}$)) from $\pmb{\alpha}$, $\pmb{\beta}$ = $\pmb{\alpha}$ - $\overline{\pmb{\alpha}}$\;I. $n_{\textnormal{BE}}(\omega)$ (1 - e$^{-\frac{\hbar \omega}{k_{B}T}}$) is the Bose Einstein (BE) factor incorporated to study the low frequency region of the Raman spectra.


\begin{figure*}[]
\centering
\includegraphics[width = 16.0 cm,angle =0]{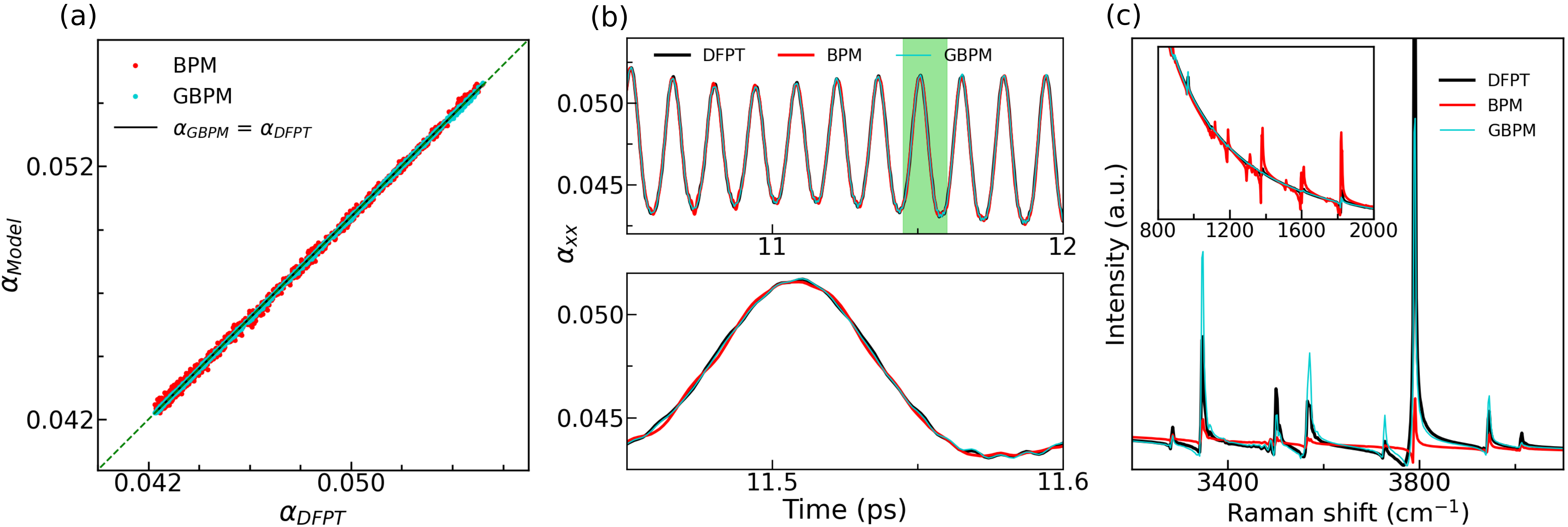}
\caption{NH$_{3}$ (a) $xx$ component of $\alpha_{BPM}$ (red circle) and $\alpha_{GBPM}$ (cyan circle) order vs $\alpha_{DFPT}$. Linear fit of GBPM order is shown in solid black line. Green dotted line represents $\alpha_{Model}$ = $\alpha_{DFPT}$. (b) Top panel: Comparison of the trajectory of $\alpha_{xx}$ from DFPT (black), $\alpha_{BPM}$ (red) and $\alpha_{GBPM}$ (cyan) order. Bottom Panel: Magnified view of the region shaded in filled green in the first panel. (c) Raman spectra from DFPT (black), BPM (red) and GBPM (cyan). Inset shows the results in the lower-wavenumber region.}
\label{Fig3} 
\end{figure*}

\section{RESULTS AND DISCUSSION}
\begin{figure*}[]
\centering
\includegraphics[width = 16.0 cm,angle =0]{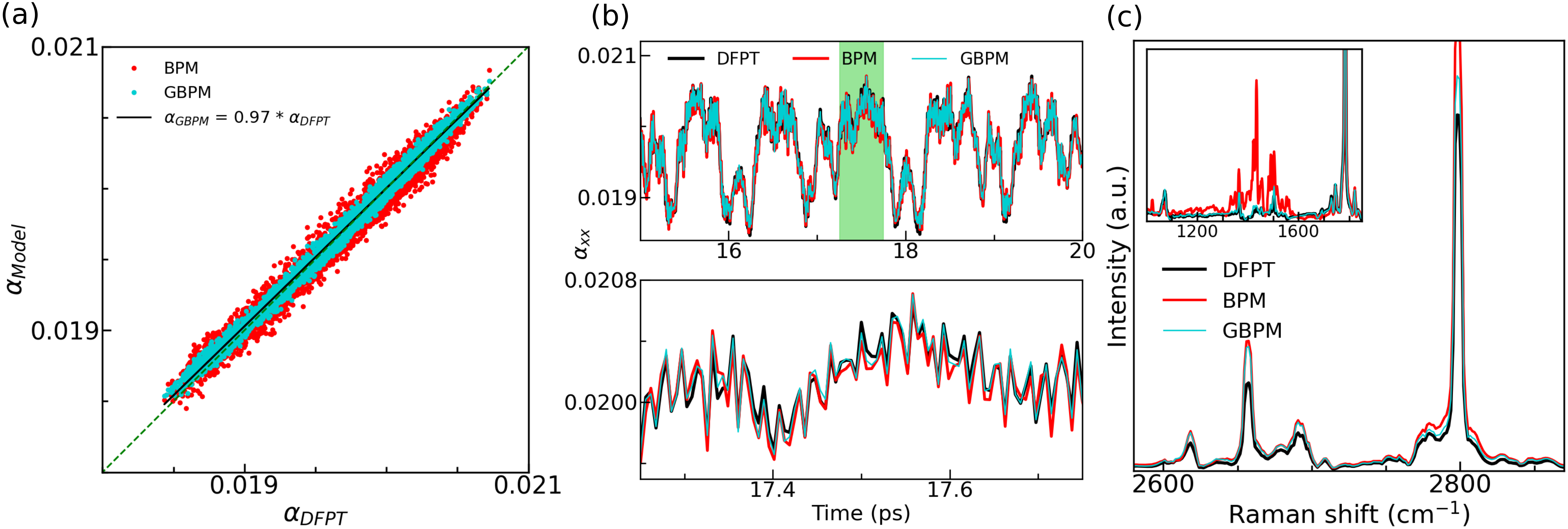}
\caption{CH$_{3}$OH (a) $xx$ component of $\alpha_{BPM}$ (red circle) and $\alpha_{GBPM}$ (cyan circle) order vs $\alpha_{DFPT}$. Linear fit of GBPM order is shown in solid black line. Green dotted line represents $\alpha_{Model}$ = $\alpha_{DFPT}$. (b) Top panel: Comparison of the trajectory of $\alpha_{xx}$ from DFPT (black), $\alpha_{BPM}$ (red) and $\alpha_{GBPM}$ (cyan) order. Bottom Panel: Magnified view of the region shaded in filled green in the first panel. (c) Raman spectra from DFPT (black), BPM (red) and GBPM (cyan). Inset shows the results in the lower-wavenumber region.}
\label{Fig4} 
\end{figure*}

\subsection{Nonlinear molecules}
To examine the improvement provided by GBPM over BPM, we first consider  H$_{2}$O which is perhaps the most important molecule in chemistry and biology. Fig.~\ref{Fig2} (a) compares the $xx$ component of $\alpha$ calculated using  BPM and GBPM with the values calculated using DFPT. It is observed that BPM-calculated values of $\alpha_{xx}$ plotted versus the DFPT values show strong scatter  and form a triangular-like distribution.  A similar triangular distribution is also observed in the plot of the polarizability calculated by the improved Thole model of H$_{2}$O molecule versus the $ab$ $initio$ values.~\cite{TholeL} The triangular shape of the distributed data in case of BPM can be attributed to the vibration  that changes the H-O-H angle. Since BPM depends on the O-H bond lengths only, it fails to take this effect into account.
By contrast, the $\alpha_{xx}$ data calculated using GBPM falls on the $y$ = $x$ line, indicating essentially perfect agreement with the DFPT results obtained using the model with only 12 adjustable parameters.

Since Raman spectra arise from polarizability fluctuations, to further examine the accuracy of the considered models for Raman spectroscopy, we compare the polarizability trajectories obtained using BPM and GBPM with that obtained using DFPT. Fig.~\ref{Fig2} (b) presents a comparison of the overall polarizability trajectory (top panel) that shows the low-frequency fluctuations as well as a magnified view of a short segment of the trajectory (bottom panel) that more clearly shows the high-frequency fluctuations.
While oscillations at similar frequencies can be observed for both DFPT and BPM trajectories, the BPM trajectory shows significant differences from the DFPT trajectory.
By contrast, the GBPM trajectory shows a perfect agreement with the DFPT results for both top and bottom panels, indicating that GBPM correctly reproduces both high- and low-frequency vibrations.  Therefore, it is expected that GBPM will generate the Raman spectra accurately compare to BPM.

The Raman spectra derived from the BPM, GBPM and DFPT-calculated polarizability trajectories are shown in Fig.~\ref{Fig2} (c) for the  high- and low-frequency regions. For the high-frequency peak, the peak shape and intensity of the BPM spectrum are similar to those of the DFPT spectrum, with some minor differences. For the GBPM spectrum, a perfect agreement with the DFPT spectrum is observed. Since both BPM and GBPM  models consider the bond-stretching vibration explicitly, they both reproduce the corresponding O-H stretching peak well.
By contrast, the low-frequency peak at 1600-2000 cm$^{-1}$ shows strong differences between the DFPT spectrum and the BPM spectrum.  This  frequency range corresponds to the vibrations that change the H-O-H angle for which the polarizability changes are not treated appropriately by BPM. It is to be noted that this H-O-H bending mode is important for probing the hydrogen bond structure of aqueous systems~\cite{jpclett_0c01259}.
By contrast, the polarizability fluctuation due to angular vibration and the corresponding changes in the response of the electron cloud between the two H atoms (see Fig.~\ref{Fig1} (a))  is included in GBPM, allowing it to fully reproduce the DFPT spectrum. Examination of the results for other triatomic-nonlinear molecules SO$_{2}$ and  H$_{2}$S shows similar trends to those discussed above for H$_2$O (see Supplementary Material (SM)). 

\begin{figure*}[]
\centering
\includegraphics[width = 16.0 cm,angle =0]{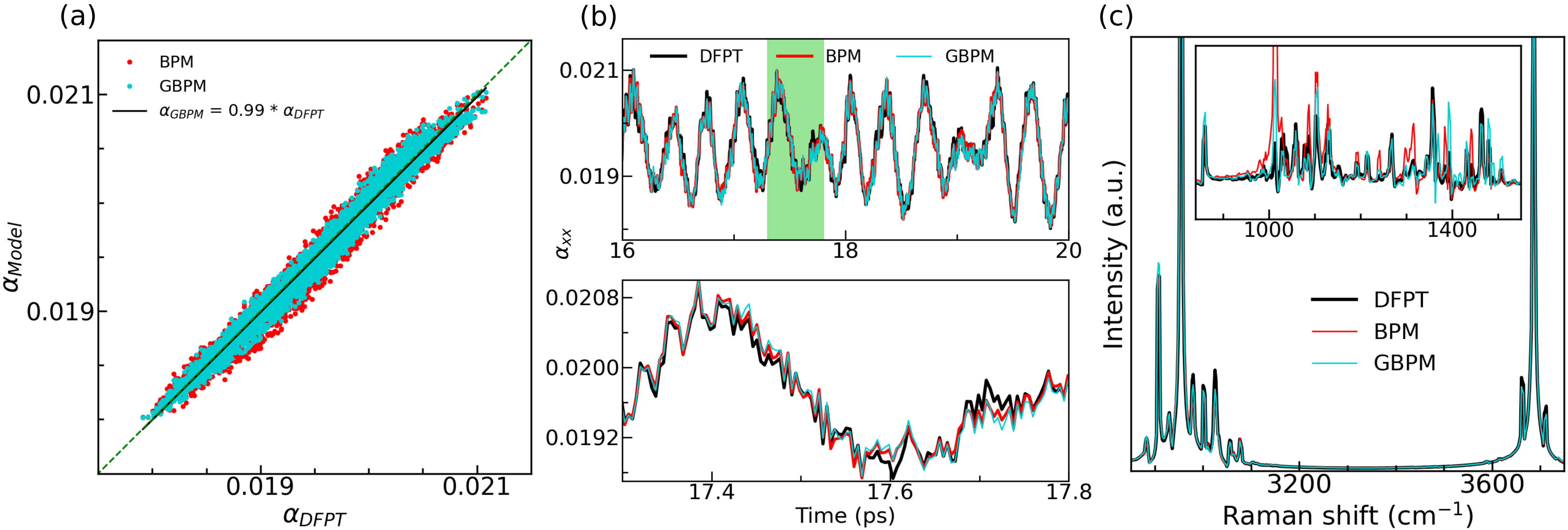}
\caption{CH$_{3}$CH$_{2}$OH (a) $xx$ component of $\alpha_{BPM}$ (red circle) and $\alpha_{GBPM}$ (cyan circle) order vs $\alpha_{DFPT}$. Linear fit of GBPM order is shown in solid black line. Green dotted line represents $\alpha_{Model}$ = $\alpha_{DFPT}$. (b) Top panel: Comparison of the trajectory of $\alpha_{xx}$ from DFPT (black), $\alpha_{BPM}$ (red) and $\alpha_{GBPM}$ (cyan) order. Bottom Panel: Magnified view of the region shaded in filled green in the first panel. (c) Raman spectra from DFPT (black), BPM (red) and GBPM (cyan). Inset shows the results in the lower-wavenumber region.}
\label{Fig5} 
\end{figure*}

\begin{figure*}[]
\centering
\includegraphics[width = 16.0 cm,angle =0]{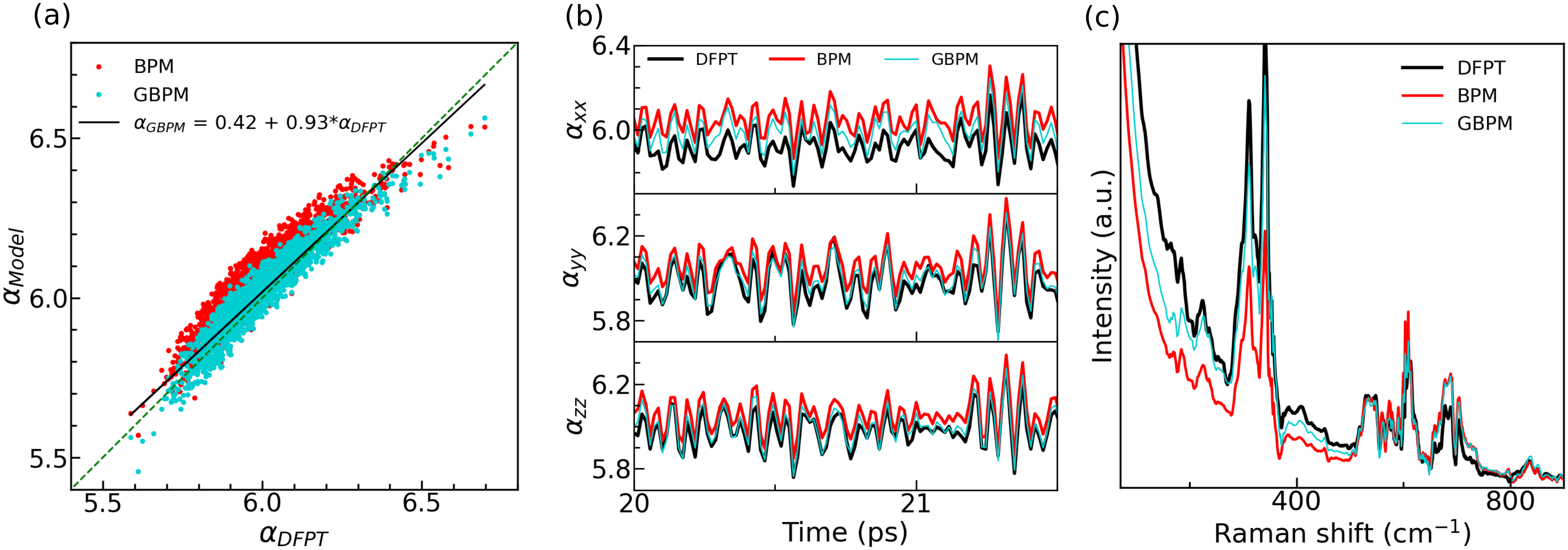}
\caption{BTO (10 atom cell) at 300 K: (Color online) (a) $xx$ component of $\alpha_{BPM}$ (red circle) and $\alpha_{GBPM}$ (cyan circle) order vs $\alpha_{DFPT}$. Linear fit of GBPM order is shown in solid black line. Green dotted line represents $\alpha_{Model}$ = $\alpha_{DFPT}$. (b) Comparison of the trajectory of $\alpha_{xx}$ (top panel), $\alpha_{yy}$ (middle panel) and $\alpha_{zz}$ (bottom panel) calculated from DFPT (black), BPM (red) and GBPM (cyan), (c) Raman spectra generated using  DFPT (black), BPM (red) and GBPM (cyan).}
\label{Fig7}
\end{figure*}

\begin{figure}[]
\centering
\includegraphics[width = 8.5 cm,angle =0]{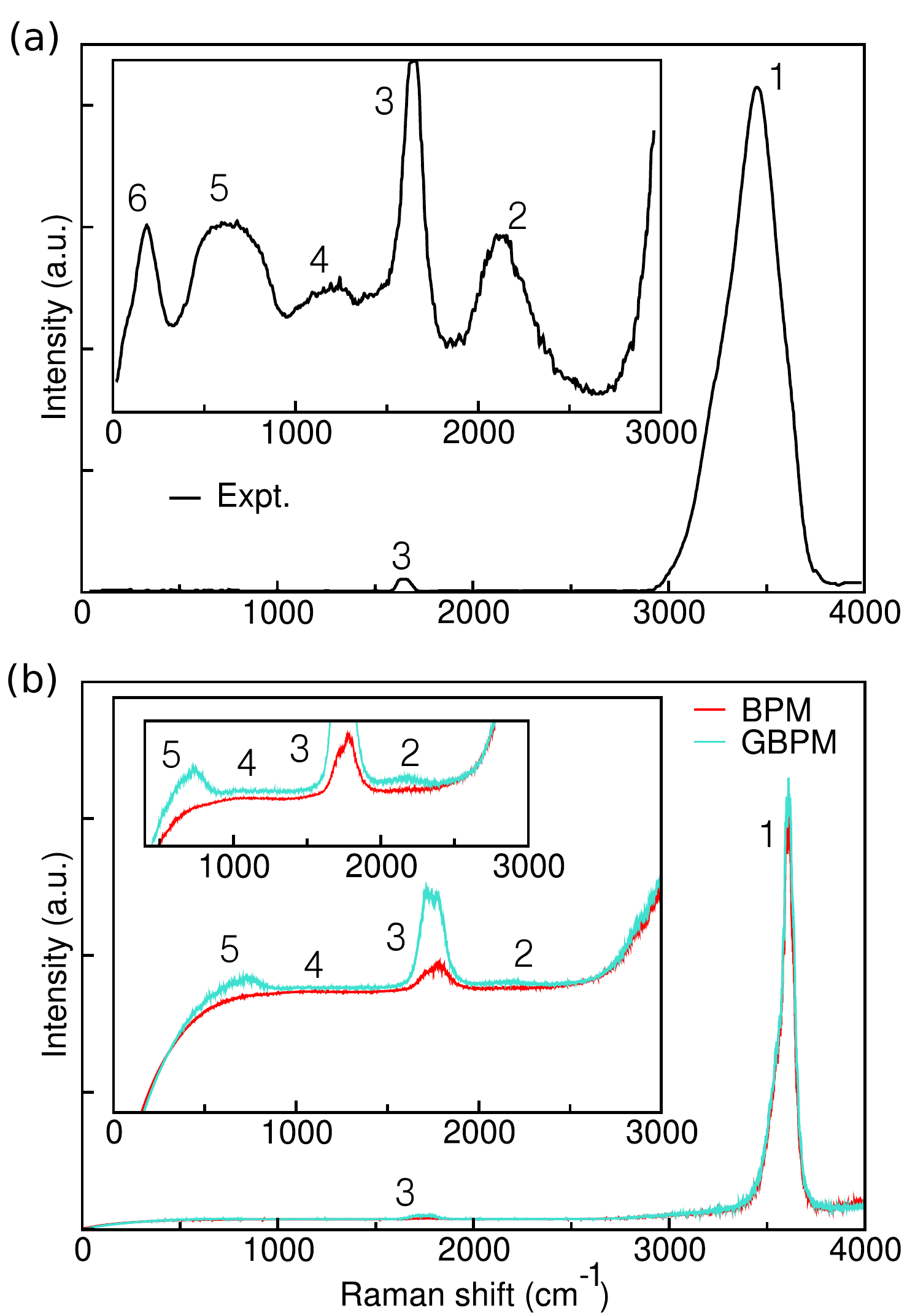}
\caption{Anisotropic Raman spectra (reduced) of liquid water (a) from experiment (Expt.) as described in Ref.~\cite{expt_Raman_aniso}, (b) using AMOEBA/BPM (red) and AMOEBA/GBPM (cyan). Inset figures show the magnified view of the lower wavenumber regions. All the peaks are marked using numbers. }
\label{Fig6} 
\end{figure}

Next, we examine the effectiveness of GBPM for  NH$_{3}$ and CH$_{4}$ which are nonlinear molecules with multiple angles.  Because similar results are obtained for NH$_{3}$ and CH$_{4}$, we only present the  NH$_{3}$ results here with the results for CH$_{4}$ shown in the SM.
We plot the values of $\alpha_{xx}$ for NH$_{3}$ calculated using both of these methods and using DFPT in Fig.~\ref{Fig3} (a). The $\alpha_{xx}$ data distribution of BPM versus DFPT shows  nearly linear trend with a small deviation of BPM from DFPT values. The triangular-shape like data distribution of the BPM is also observed for NH$_{3}$ but with a much smaller area than for water (see SM). As discussed in our previous work,~\cite{atanu_BPM} this is due to the fact that for most angle vibrations in NH$_{3}$, the  effect of the change in the bond angle on polarizability is proportional to the square of the change in the bond angle ($\delta \theta^2$), in contrast to the linear effect proportional to $\delta \theta$ for water. A better agreement with DFPT results is observed for the GBPM model and the triangular-shape like data distribution is completely removed. In the case of CH$_{4}$, similar good agreement with DFPT results is obtained for the GBPM  model (see SM).

For better comparison of the models, we plotted the time-dependent polarizability trajectories of NH$_{3}$ obtained from BPM, GBPM and DFPT in Fig.~\ref{Fig3} (b). Clearly, BPM shows overall agreement with the DFPT trajectory; however, as observed from the magnified view focusing on $t$ = 11.5 ps region (Fig.~\ref{Fig3} (b)), the high-frequency fluctuations of the DFPT trajectory are not reproduced by the BPM.  By contrast, GBPM-calculated polarizability trajectory  shows an almost perfect agreement with the DFPT results. The Raman spectra obtained from these trajectories are shown in Fig.~\ref{Fig3} (c). The high frequency peak positions from DFPT-calculated spectra are well reproduced by GBPM and BPM, but only GBPM reproduces the DFPT peak intensities. Furthermore,  BPM does not reproduce the low-frequency DFPT peak positions and intensities and even exhibits additional peaks that are absent in the DFPT spectrum.


Next, we examine the utility of GBPM for   CH$_{3}$OH which can be considered to be a combination of H$_2$O and CH$_4$, contains the C-O-H bond and has a much more complex spectrum. We expect that  GBPM should exhibit similar level of accuracy for this system to that for H$_2$O and CH$_4$.  In Fig.~\ref{Fig4} (a), we plot $\alpha_{xx}$ calculated from BPM and GBPM (here GBPM contains the terms in BPM and the H-H, H-C and H-O terms of the  H-C-H, H-O-C and H-C-O  angles, respectively)
versus the DFPT-calculated values. Both the  BPM and GBPM data fall around the $y$ = $x$ line but a clear decrease in the scatter is observed for  the GBPM results relative to the  BPM data. We then compare the $\alpha_{xx}$ trajectories obtained using BPM, GBPM and DFPT in Fig.~\ref{Fig4} (b). The polarizability trajectory obtained using BPM reproduce the overall fluctuations of the DFPT trajectory  (see top panel of Fig.~\ref{Fig4} (b)), but does not reproduce the fine features of the high-frequency fluctuations  of the DFPT-calculated polarizability trajectory (see bottom panel of Fig.~\ref{Fig4} (b)). The GBPM-calculated polarizability trajectory shows a better agreement with the DFPT results both overall and for the fine features of high-frequency fluctuations. These observations are reflected in the Raman spectra plot as shown in Fig.~\ref{Fig4} (c). Both BPM and GBPM reproduce the peak positions and shapes of the DFPT-calculated Raman spectra  in the high-frequency region.
However, BPM shows differences from the DFPT spectrum in the low-frequency region; in particular, the BPM spectrum strongly overestimates the Raman intensity in the 1200-1600~cm$^{-1}$ range.
The better agreement of the GBPM results in the low-frequency region is due to fact that GBPM takes into account  the low-frequency angular vibration by explicitly considering the H-H, H-C and H-O interactions of the  H-C-H, H-O-C and H-C-O angles, respectively. To check further the impact of each of these interactions on GBPM accuracy, we compared the calculated spectra considering different angular contribution (see SM). The results suggest that the  H-C-H and  H-O-C angular interactions  are important for capturing the low-frequency polarizability fluctuations, whereas the H-C-O angle is less important.

Proceeding to more complex systems, we applied BPM and GBPM to CH$_{3}$CH$_{2}$OH which can be considered to be a combination of CH$_{4}$ and CH$_{3}$OH, and which contains C-C-H and C-C-O bond angles absent in the smaller molecules discussed above. Fig.~\ref{Fig5} (a) compares the $\alpha_{xx}$ calculated using BPM and GBPM plotted versus the DFPT values.
Both BPM and GBPM results are scattered around the $y$=$x$ line showing overall agreement with the DFPT-calculated $\alpha_{xx}$ values. However, the scatter is smaller for GBPM results indicating better agreement with DFPT $\alpha_{xx}$ values.
The BPM, GBPM and DFPT polarizability trajectories are compared in Fig.~\ref{Fig5} (b).
Both BPM and GBPM reproduce the overall trend of the DFPT-calculated polarizability fluctuations, but neither model shows a perfect agreement with the DFPT results.
Examination of the Raman spectra presented in Fig. ~\ref{Fig5} (c) shows good agreement between the BPM-, GBPM- and the DFPT-calculated spectra for the high-frequency peaks.  However, BPM shows differences from the DFPT spectrum in the low-frequency region; in particular, the BPM spectrum strongly underestimates the Raman intensity in the 1200-1600~cm$^{-1}$ range. This range is consistent with the frequency range of the large differences between the GBPM and BPM Raman spectra of CH$_3$OH.


\subsection{Condensed Matter systems - Perovskite Oxide and  Liquid water}

To demonstrate the advantage of GBPM for simulations of Raman spectra of condensed matter systems, we investigate  BTO ferroelectric complex oxide and liquid water.  These systems provide a more challenging test for atomistic models due to their more complex atomic arrangements, and the possibility of low-energy distortions (e.g. ferroelectric Ti off-center displacements in BTO and hydrogen bond formation and breaking for liquid water) that lead to greater variations in bond lengths and bond angles compared to gas-phase molecules. 

For BTO, we compare the DFPT, BPM and GBPM results for a 10-atom cell in Fig.~\ref{Fig7}. While larger cells (e.g. 40 or more atoms) provide a better representation of BTO dynamics and thus a better test of the models, DFPT calculations for a trajectory of a larger cell are much more computationally expensive. Therefore, we chose the 10-atom cell to demonstrate the improved accuracy of GBPM relative to BPM applied to the solid-state BTO.

Here, in addition to the Ti-O bonds, we defined parameters for the O-O interactions in GBPM.
We note that O-O interactions account for the effect of the changes in the interactions between the orbitals of these O atoms in extended electronic states of BTO at the top of the valence band, rather than long-range non-bonding interactions such as dispersion and electrostatic forces.  The plot of the maximally localized Wannier function of the O-$p$ orbital in rhombohedral BTO presented in  
Fig. S9 (See SM) shows that there are states that connect the nearest-neighbor O atoms forming a 90$^\circ$ O-Ti-O angle (O atoms O1 and O2 in the figure). Additionally there is significant charge density located along the line connecting these  two O atoms. These states are due to the lone pair O orbitals which provide the top of the valence band of BTO and thus are important for polarizability.  It is clear that changes in the  distance between O atoms O1 and O2 will affect the shape of the charge density and therefore the polarizability of the BTO.  This is due to the changes in the quantum mechanical orbital overlap  interactions rather than long-range forces.
Ba-O and Ba-Ti interactions are omitted from BPM and GBPM because Ba motion contributes little to the change in  $\alpha$ during the MD trajectory.~\cite{atanu_BPM}

It is observed that GBPM $\alpha_{xx}$ values plotted versus DFPT $\alpha_{xx}$ show less scatter around the $y$ = $x$ line   the BPM $\alpha_{xx}$ values. However, both BPM and GBPM show some deviations from the $y$ = $x$ line for large and small values of $\alpha_{xx}$. Examination of the polarizability trajectories shows that while neither GBPM nor BPM perfectly reproduces the DFPT trajectory, the GBPM trajectory is significantly closer to the DFPT trajectory. Comparison of the Raman spectra shows that both BPM and GBPM reproduce the peak positions of the DFPT spectrum; however, BPM obtains incorrect relative intensities of the low-frequency peaks ($<500$cm$^{-1}$), and also shows greater broadening for the peak at 650 cm$^{-1}$.
GBPM shows improvement to essentially perfect agreement with the DFPT spectrum peak intensity for the peaks at 210, 360, and 600 cm$^{-1}$,
some degree of improvement for the features at $<200$cm$^{-1}$, 350 cm$^{-1}$, and 400-430 cm$^{-1}$, 
and no improvement for the peaks at 700 and 720 cm$^{-1}$.  Thus, while significant improvements are obtained by including the O-O interactions and thus the effects of variation in the O-Ti-O angles on $\alpha$, some other effects are still not accounted for in GBPM, leading to imperfect agreement between the GBPM and DFPT spectra.

The evaluation of the effectiveness of the polarizability model for the calculation of Raman spectrum of liquid water is more complex due to the lack of DFPT reference information. Molecular dynamics simulations of Liquid water require the use of fairly large cells (more than 200 molecules) and simulations times on the order of 0.1-1~ns to capture the complex dynamics of hydrogen bond breaking and formation as well as a small simulation time step of 0.2~fs to property represent the high-frequency oscillations of O-H bonds. This makes the use of first-principles methods quite expensive  for obtaining the MD trajectory. Even more importantly, the currently implemented DFPT methods cannot be applied to such a large system. Thus, no DFPT reference data for the polarizability trajectory and the derived Raman spectrum are available for direct comparison with the BPM and GBPM atomistic polarizabilty models. Since typical simulations of water use atomistic model and seek to reproduce experimental behavior of water, we use atomistic MD simulations of liquid water obtained with the AMOEBA potential to obtain the liquid water trajectory and then apply our BPM and GBPM models of polarizability to the structures along the trajectory to obtain the Raman spectrum that is then compared to experimental Raman spectrum.
  We choose the AMOEBA potential because it shows good agreement with the highly accurate and computationally expensive MB-pol potential for several key properties of liquid water  while having a much lower computational cost.~\cite{AMOEBA_MBpol,MBpol_water,MBpol_new}
   In this case, the discrepancies between the theoretical and experimental Rama spectra can be due to the deficiencies of the AMOEBA atomistic potential and/or due to the deficiencies in the representation of the true polarizability of the liquid water system by the BPM and GBPM polarizability models. While it is difficult in some cases to separate the two possible sources of error, comparison to experimental spectrum still provide valuable information regarding the accuracy of the theoretical method.

Figs.~\ref{Fig6} (a) and (b) present the reduced anisotropic Raman spectra of liquid water at 300 K obtained from experiment,~\cite{expt_Raman_aniso} and those derived using the BPM and GBPM models of the polarizability of the H$_{2}$O molecule and trajectories generated by MD simulations with the AMOEBA potential. We refer to these MD-derived spectra as the AMOEBA/BPM and AMOEBA/GBPM spectra, respectively.

An examination of the overall spectra shows that in agreement with the experimental spectrum, the AMOEBA/BPM and AMOEBA/GBPM spectra show one very intense peak at high frequency  with some weak features at lower energies (see Fig.~\ref{Fig6}). The high-frequency peak (peak 1) in the  AMOEBA/BPM and AMOEBA/GBPM spectra  is  blueshifted by 160 cm$^{-1}$ from the experimental peak at 3450 cm$^{-1}$. 
This discrepancy is due the limitation of the classical  liquid water simulations using the AMOEBA force field and is in agreement with previous work that showed that the neglect of quantum effects in MD simulations of liquid water leads to a blueshift of the vibration spectrum by 170~cm$^{-1}$.~\cite{acs.jpca.0c05557}

Examination of the lower frequencies shows that the peak at 2100 cm$^{-1}$ (peak 2) is entirely absent in the AMOEBA/BPM spectrum and is very faint in the AMOEBA/GBPM spectrum. This is due to the inability of the atomistic potentials to accurately reproduce the Fermi resonance between the bend overtone and the OH stretch modes that gives rise to this peak, as has been previously discussed by Medders et al.~\cite{MBpol_water_Raman}  The H-O-H bend peak at 1650 cm$^{-1}$ (peak 3) is present in both BPM and GBPM spectra, but the GBPM spectrum gives a more accurate shape of this peak and also has higher intensity, showing better agreement with experimental results.
Specifically,  the experimental H-O-H bend peak has an almost symmetric shape, with the rise on the left side that is slightly more rapid  than the decay on the right side. However, the AMOEBA/BPM peak appears much more skewed to the right, with the rise on the left side slower than the decay on the right side. By contrast, for the AMOEBA/GBPM peak, the rise on the left side is slightly more rapid than the decay on the right side, in agreement with the experimental results. Thus, the shape of the AMOEBA/GBPM H-O-H bend peak is more consistent with the experimental results than that of the AMOEBA/BPM peak.

For the peak inensity, comparison of the experimental O-H stretch and H-O-H bend (Fig. ~\ref{Fig6} (a)) peaks shows an intensity ratio between the peaks of 36:1.  The values of this ratio obtained for the AMOEBA/BPM and AMOEBA/GBPM spectra are 382:1 and 107:1, respectively.  Thus, the intensity of the H-O-H bend (relative to the O-H) stretch is underestimated in the theoretical spectra. However, this underestimation is worse for AMOEBA/BPM than for AMOEBA/GBPM. Thus, the higher intensity of the H-O-H bend in the AMOEBA/GBPM is closer to the experimental results.

Since both AMOEBA/BPM and AMOEBA/GBPM spectra use the same AMOEBA trajectories, since AMOEBA is known to be a good potential for liquid water, we attribute the better agreement with the experimental shape and intensity of the AMOEBA/GBPM spectrum to the greater accuracy of the GBPM model for the H-O-H bend of the monomer as shown in  Fig. ~\ref{Fig6} of the paper.

Comparison of the MD-derived and experimental spectra shows that none of the experimental low-frequency peaks at  200 (peak 6) cm$^{-1}$, 650 cm$^{-1}$ (peak 5) and 1200 cm$^{-1}$ (peak 4) are reproduced in the AMOEBA/BPM spectrum. The peak at 200 cm$^{-1}$ is entirely absent, and the peaks at 650 cm$^{-1}$ and 1200 cm$^{-1}$ are at best very faint.  By contrast, the AMOEBA/GBPM spectrum shows a pronounced broad peak at 650 cm$^{-1}$ and a weak broad peak at 1200 cm$^{-1}$, similar to the experimental spectrum. However, the peak at 200 cm$^{-1}$ which is related to the vibrations of hydrogen bonds is  not reproduced in the AMOEBA/GBPM spectrum, most likely because GBPM does not take  intermolecular  interactions into account.  Thus, while the AMOEBA/GBPM spectrum does not fully reproduce the experimental spectrum, it achieves considerably improved qualitative agreement compared to the AMOEBA/BPM spectrum.

Our results suggest that the  high accuracy of BPM for the monomer enables qualitative agreement of the spectrum even for liquid water due to the correct modeling of the effects of the  geometrical changes of the individual molecules on polarizability.  While it is difficult to draw unambiguous conclusions from the comparison of model spectrum to the experimental spectrum due to the convolution of the effects of the error of the AMOEBA force-field and our  polarizability model,it is logical to expect that an improvement in the modeling of the intra-molecular geometry effect on polarizability will lead to improved agreement even in systems where inter-molecular interactions are important. Our results support this expectation and tentatively suggest that GBPM can qualitatively reproduce some features of the liquid water spectrum. 
This  is in agreement with previous work that demonstrated the importance of the H-O-H bending mode for accurate representation of the  vibrational properties of liquid water.~\cite{jpclett_0c01259}

While GBPM fully reproduces the DFPT results fo gas molecules, it shows deviations from the DFPT trajectories and spectrum for BaTiO$_3$ and also does not fully reproduce the experimental Raman spectrum of liquid water. Thus, GBPM shows worse accuracy for condensed-matter systems compared to gas-phase molecules.  This is due to the importance of non-covalent intermolecular interactions for condensed-matter systems. For example, for a liquid, the electronic wavefunction of the system is a function of atomic coordinates of all of the atoms in the system, which means that the electronic wavefunction of each individual molecule in a liquid is affected by the positions of the atoms in all other molecules. Therefore, the electronic polarizability $\alpha$  of a given molecule is also affected by the atomic positions in all other molecules.  Thus, treating the polarizability of a liquid as a sum of polarizabilities of the individual gas phase molecules as is done in GBPM and BPM will introduce some error.
In chemical language, for a given molecule in the liquid, the response of its electron cloud to the applied electric field will not only be affected by its geometry but also by the effects of the electron clouds of other molecules. Such effects are likely to be particularly important in liquid water which has relatively strong hydrogen-bond intermolecular interactions. Therefore, it is expected that the lack of treatment of the effects of intermolecular interactions in the BPM and GBPM models which consider the polarizability of each molecule to be functions of the coordinates of that molecule only will lead to worse accuracy for liquid than for a gas phase.
A similar argument can be made for the BaTiO$_3$ system where non-bonded interactions and long-range effects will influence the polarizability so that a fully local model such as BPM and GBPM  will inevitably introduce some error.

\section{CONCLUSIONS}
We have shown that the bond polarizability model for calculations of Raman spectra from MD simulations can be substantially improved by generalizing it to include the interactions between atoms considered to be non-bonding, for example the interaction between the two H atoms in H$_2$O.  We find that the inclusion of these terms leads to perfect agreement between the model and DFPT polarizability values for water and other molecules.  This is due to the non-negligible wavefunction density between such non-bonding atom pairs in the frontier orbitals that play the dominant role in determining the electronic polarizibility.  We find that for the more complex liquid water and solid-state BTO systems, GBPM achieves a substantial improvement over BPM in terms of the agreement between the model and DFPT polarizability and Raman spectra. For liquid water, GBPM qualitatively reproduces all of the experimental Raman peaks except for the low-frequency peak at 200 cm$^{-1}$ related to hydrogen bonding. By contrast, the BPM Raman spectrum is qualitatively less correct compared to GBPM. Nevertheless, the lack of the hydrogen bonding peak point to the need to further improve the polarizability model for water to include the effects of intermolecular interactions on the polarizability, either by including additional analytical terms to account for intermolecular interactions or by parameterizing machine learning models for the intermolecular interactions only.  Similarly, for BaTiO$_{3}$, the agreement between GBPM and DFPT Raman spectra derived from MD trajectory of 10-atom supercell is not perfect, pointing to the need to include additional interactions in the polarizability model.  We hope that this study will motivate further efforts in the development of atomistic models of electronic polarizability and the use of MD simulations for the interpretation of Raman spectroscopy results.



\section{SUPPLEMENTARY MATERIAL}
The supplementary material file contains the results of other molecules  SO$_{2}$, H$_{2}$S, CH$_{4}$ and NH$_{3}$, $R^{2}$ value (coefficient of determination) of BPM and GBPM for the studied systems, accuracy test for different level of GBPM in case of CH$_{3}$OH, convergence test with number of structures for H$_{2}$O and
CH$_{3}$OH, comparison of first and second order GBPM for H$_{2}$O, justification of the requirement of O-O non-bonded interaction in BTO.

\begin{acknowledgments}
A.P., N.R.S., and I.G. acknowledge the support of the Army Research Office
under Grant W911NF-21-1-0126 and Army/ARL via the
Collaborative for Hierarchical Agile and Responsive Materials (CHARM) under cooperative
agreement W911NF-19-2-0119. A.P., N.R.S. and I.G acknowledge additional support from Israel Science Foundation under Grant 1479/21. 
\end{acknowledgments}

\section*{Data Availability}
The data used in this work are available from the corresponding author upon reasonable request.

\bibliographystyle{apsrev4-1}
\bibliography{sbiblio.bib}

\end{document}